
\magnification=\magstep2
\def\para{\par\noindent}
\def\sqr#1#2{{\vcenter{\vbox{\hrule height.#2pt
        \hbox{\vrule width.#2pt height#1pt \kern#1pt
          \vrule width.#2pt}
        \hrule height.#2pt}}}}
\def\square{\mathchoice\sqr34\sqr34\sqr{2.1}3\sqr{1.5}3}
\newcount\notenumber

\def\note{\advance\notenumber by 1
\footnote{$^{\the\notenumber}$}}
\baselineskip 20pt
\centerline{{\bf Anomalously slow relaxation in the diluted Ising model
 below}}
\centerline{{\bf the percolation threshold}}\para
\vskip 0.25cm
 \para S.~Jain,
 \para School of Mathematics and Computing,\para
 University of Derby,\para
 Kedleston Road,\para
 Derby DE22 1GB,\para
 U.K.\para
\vskip 0.25cm
\para E-mail: S.Jain@derby.ac.uk
\vskip 2.0cm
Classification Numbers: 0520, 0550, 7510H, 7540D
\vskip 0.5cm
$$\vbox{\settabs 7\columns
\+Keywords: &Disorder, Dynamics, Dynamic Phase
 Transition, Ising,\cr
\+\quad &Monte Carlo, Percolation\cr}$$
\vskip 0.5cm
Running title: Anomalous relaxation below the percolation threshold
\vskip 2.0cm
\vfill\eject
\para ABSTRACT
\par The relaxational behaviour of the bond-diluted two-dimensional
 Ising model below
the percolation threshold is studied using Monte Carlo techniques.
The non-equilibrium decay of the
 magnetization, M(t), and the relaxation of
the equilibrium spin-spin autocorrelation function, C(t), are monitored.
\par The behaviour of both $C(t)$ and $M(t)$ is found to satisfy
 the Kohlrausch law of a stretched exponential with the same
 temperature-dependent exponent. The Kohlrausch exponent does not appear to
depend on the bond concentration.
\par The results indicate that we
 are not yet in the asymptotic regime, even when
C(t) and M(t) are less than $10^{-4}$.
\vskip 0.25cm
\vfill\eject
\para 1.\ INTRODUCTION
\par The dynamics of disordered magnetic systems such as the dilute
ferromagnet have been
studied recently both by analytic techniques [1-3] and Monte Carlo
 simulations [4-7].
\par In a ferromagnet [8], with a concentration $p$ of
 sites or bonds, $T_c(p)$ denotes
the critical temperature below which the system exhibits magnetic long-range
order; this regime is indicated by $F$ in figure 1 which depicts a schematic
phase diagram for a diluted ferromagnet. Above $T_c(p)$ the system is a
paramagnet ($P$). The boundary separating the two regimes intersects the
zero-temperature axis at $p_c$, the percolation threshold, i.e. $T_c(p=p_c)=0$.
\par The highest possible critical temperature for ferromagnetic
 order in the
disordered system permitted by a rare statistical fluctuation of the disorder
over the entire system is called [9] the \lq Griffiths
 temperature\rq, $T_G$. For
a diluted ferromagnet $T_G$ is the transition temperature of the pure system,
i.e. $T_G\equiv T_c(p=1)$. The regime $T_c(p) < T < T_G$ is usually referred
 to as the \lq Griffiths phase\rq\ ($G$) [10].
\par The Hamiltonian of the dilute two-dimensional Ising ferromagnet
 is given by
$${\it H} = -\sum_{<i,j>} J_{ij} S_iS_j\eqno(1)$$
where $S_i=\pm 1$ are Ising spins, the summation runs over nearest neighbours
only and the exchange interactions are chosen from
$$P(J_{ij}) = (1-p)\delta (J_{ij}) + p\delta (J_{ij}-1).\eqno(2)$$
Setting both Boltzmann's constant and the exchange coupling to unity implies
that $T_G=2/\ln (1+\sqrt(2))$ for this system.
\par A key quantity of interest in the study of the dynamics of the diluted
ferromagnet is the equilibrium spin-spin autocorrelation function $C(t)$,
 defined by
$$C(t) = [<S_i(t)S_i(0)>],\eqno(3)$$
where $<\dots >$ and $[\ldots ]$ denote thermal and disorder
 averages, respectively.
\para The behaviour of $C(t)$ depends on the phase under question.
A novel form of dynamic scaling that involves $t/\xi^{2+Z}$ results from an
assumption [1] that the dynamics are dominated by compact clusters; here $\xi$
is the correlation length of order parameter fluctuations in the {\it pure}
$(p=1)$ system and $Z$, the dynamical critical exponent, is also that of the
{\it non-dilute} system.
\par In this paper we present new data for the diluted two-dimensional Ising
model
{\it below} the percolation threshold. Our main objective is to try and
establish
the validity of the predicted asymptotic behaviour of the equilibrium spin-spin
autocorrelation function.
\par In section 2 we give a brief review of the theory and the predicted
asymptotic form of $C(t)$. The simulation procedure is outlined in section 3
and
is followed by our results in section 4. The conclusion is given in section 5.
\para 2.\ THEORY
\par We review the theory which is primarily due to Bray [1,3]. In the
following we shall concern ourselves with exponential factors only and ignore
all power-law prefactors. Although the theory
 refers to site-dilution, the results are also applicable to bond-dilution with
minor modifications.
\par In Bray's [1] approach one writes
$$ C(t) = \sum_{\alpha } n_{\alpha } e^{-t/\tau_{\alpha }},\eqno(4)$$
where $n_{\alpha }$ is the probability that a given site belongs to a cluster
of type $\alpha $ (here $\alpha $ refers to size, shape and composition) and
$\tau_{\alpha }$ is the relaxation time for the cluster.
Two distinct regimes are predicted.
\par In the paramagnetic phase $T\ge T_G$, clusters of fully occupied sites
(or bonds) dominate the long-time dynamics and a straight-forward variational
argument leads to
$$\ln C(t)\sim \cases{-t^{2/(2+Z)}&for $t \ll \xi^{2+Z}$\cr
                      -t&for $t \gg \xi^{2+Z}$\cr}\eqno(5)$$
 Note that for $T=T_G$ it is expected that $\ln C(t)\sim -t^{2/(2+Z)}$
for all $t$ since the correlation
 length (of the pure system) is
then infinite.
\par In the Griffiths phase $(T_c(p) < T < T_G)$ the behaviour depends
on the form of the dynamics and it will be assumed that we are dealing with
\lq model A\rq\ i.e. non-conserved dynamics [11].
$C(t)$ is bounded below [2] by any subset of the terms forming the
 sum in equation (4). In particular,
$$C(t)\ge \sum_{l} n_{l} e^{-t/\tau_{l}},\eqno(6)$$
where now $l$ specifies just the cluster size. The probability that a given
site belongs to a cluster of \lq volume\rq\ $l^2$ is given to leading order
by
$$n_{l}\sim p^{l^2} = exp{[-l^2\ln (1/p)]}.\eqno(7)$$
The relaxation time, $\tau_{l}$, is given by the Arrhenius result for the
global reversal of the magnetization. Thus,
$$\tau_{l}\sim exp[\sigma l/T],\eqno(8)$$
where $\sigma $ is the bulk surface tension at temperature $T$. It follows from
equations (6) and (7) that
$$C(t)\ge\sum_{l} exp{[-l^2\ln (1/p) - t/\tau_l]}.\eqno(9)$$
The summation can be performed by substituting for $\tau_{l}$ from equation (8)
and then using the method of steepest descent for large $t$. As a result
$$-\ln C(t)/(\ln t)^2\le A_{max}\quad {\rm for}\ t\gg\xi^{2+Z}\eqno(10)$$
with
$$ A_{max} = \bigl(T/\sigma\bigr)^2\ln (1/p).\eqno(11)$$
\par Below the percolation threshold most clusters are ramified. The relaxation
time,
$\tau_x$, of a ramified cluster containing $x$ sites is bounded above by that
of a corresponding compact cluster with the same number of sites, that is
$$\tau_x\le exp[\sigma x^{1/2}/T].\eqno(12)$$
The number of $x$ clusters per site below the percolation threshold and for
large $x$ is given by [12]
$$n_x\sim exp[-a(p) x],\eqno(13)$$
where $a(p)\rightarrow 0$ as $p\rightarrow p_c-$. It follows from equations
(4),
 (12)
and (13) that an upper bound on $C(t)$ is given by
$$C(t)\le\sum_x exp[-a(p)x - t e^{-\sigma x^{1/2}/T}].\eqno(14)$$
The summation, which can once again be performed by the method of steepest
 descent for large $t$, leads to
$$-\ln C(t)/(\ln t)^2\ge A_{min}\quad {\rm for}\ t\gg\xi^{2+Z}\eqno(15)$$
with
$$ A_{min} = \bigl(T/\sigma\bigr)^2 a(p).\eqno(16)$$
Thus, we can combine equations (10) and (15) and write
$$\ln C(t)\sim -A (\ln t)^2\quad {\rm for}\ t\gg\xi^{2+Z}\eqno(17)$$
where
$$ A_{min}\le A \le A_{max}, \eqno(18)$$
and the amplitude $A$ depends on the system parameters.
\par A consequence of
equation (17) is that the relaxation is non-exponential throughout the
Griffiths
phase. Hence, a dynamical phase transition occurs at the Griffiths temperature
from exponential to non-exponential relaxation. Furthermore, the bounds on $A$,
 equation (18), confirm that this is the correct asymptotic form below the
percolation threshold. As $a(p)$, and hence $A_{min}$, vanishes at $p_c$, the
upper bound is of no value for $p\ge p_c$. It has, however, been argued that
equation (17) holds throughout the entire Griffiths phase [1, 2, 13].
\par Computer simulations have been performed at the percolation threshold
[4, 5, 7] and both slightly above [5, 6] and below [5] $p_c$. The Monte Carlo
data have been found to satisfy the Kohlrausch [14] law of a stretched
 exponential which is given by
$$\ln C(t)\sim -B t^n\qquad\qquad 0 < n \le 1 \eqno(19)$$
where $B$ and $n$ are temperature-dependent parameters. The numerical data
can be fitted by the Kohlrausch form over a remarkable range of times, from
very
short times, up to the latest times achieved in the simulations. In particular,
the stretched exponential form has been observed for the regime $t\gg\xi^{2+Z}$
[4, 7].
\par Most of the simulation work so far has concentrated on the behaviour of
the
model at the percolation threshold. As argued above, equation
 (17) is believed to be the correct asymptotic form for the relaxation of
$C(t)$
for $p < p_c$. This asymptotic form remains to be seen in the simulations
for $p\ge p_c$. Hence, it is important to establish that equation (17) does
 indeed describe the relaxation of $C(t)$ for $p < p_c$. There are very limited
numerical results available for systems below $p_c$. The only other simulation
that we are aware of is that of Colborne and Bray [5] who studied
three-dimensional
Heisenberg and Ising models slightly below $p_c$. Good agreement with the
 theoretical predictions was found for the Heisenberg system. The data for the
Ising system, on the other hand, were not at all conclusive.
\par We, therefore, investigate the behaviour of a diluted Ising system
 {\it below} the percolation threshold. Our underlying aim is to look for
the predicted asymptotic form of $C(t)$.
\para 3.\ THE SIMULATION PROCEDURE
\par The simulations were performed on a two-dimensional bond-diluted Ising
model on a square $64\times 64$ lattice. The Hamiltonian for our model
is given by equation (1). For any given $p$ and temperature, $T$, the nearest-
neighbour interactions are chosen according to equation (2). We studied the
system
below the percolation threshold $p_c (=1/2)$ for two bond concentrations:
$p=0.1$
and $p=0.25$. We impose full periodic boundary conditions and perform
 simulations using conventional [15] Monte Carlo techniques.
\par The correlation length, $\xi (T,p)$, of the {\it diluted} system
 is given by [8]
$$\xi^{-1}(T,p) = \xi^{-1}_T(T,p_c) + \xi^{-1}_p(0,p),\eqno(20)$$
where $\xi_T$ and $\xi_p$ are the thermal and percolation correlation
 lengths, respectively.
Throughout our simulations we have [16] $min(\xi_T,\xi_p)\ll$ the linear size
of the lattice; the dominant clusters are not particularly large
at the timescales of interest. Consequently, we do not expect our results to
be influenced by finite-size effects.
\par At the beginning of the simulation all spins are aligned. The square
lattice is regarded as two inter-penetrating sublattices and all the spins of
a given sublattice are updated in parallel according to Metropolis dynamics
with
the transition probability given by
$$ W(S_i\rightarrow -S_i) = min[1, exp(-\Delta E/T)],\eqno(21)$$
where $\Delta E$ is the energy change involved. The thermal and disorder
averages
appearing in equation (3) are replaced by averages over time and samples,
 respectively.
\par In the simulations we initially monitor the non-equilibrium decay of the
magnetization, $M(t)$, defined by
$$M(t) = {1\over N}\sum_{i} S_i(t)\eqno(22)$$
and here $N=4096$. The system is assumed to be in equilibrium once the
magnetization has decayed to zero within the statistical error. We then measure
the equilibrium spin-spin autocorrelation function from
$$C(t) = {1\over N}\sum_{i} S_i(t) S_i(0),\eqno(23)$$
where $t=0$ refers to an equilibrium state.
\par At asymptotic timescales the only mode which remains to be equilibrated
is that associated with a global reversal of the magnetization. As a
 consequence,
Colborne and Bray [5] have argued that the non-equilibrium decay of $M(t)$
should exhibit the same asymptotic behaviour as the relaxation
of the equilibrium $C(t)$. The data for $M(t)$
will be examined to see whether this is indeed the case.
\par To obtain statistically reliable data we averaged over at least $10^5$
samples for each $T$ and in some cases over $10^6$ samples were used.
\para 4.\ RESULTS
\par We now discuss our results. It should be noted that in our simulations we
managed to obtain values of $M(t)$ and $C(t)$ an order of magnitude less than
those achieved in previous work [4-7].
\par Our results are summarised in figures 2 to 6. Figure 2(a) gives
 a plot of
our data for $p=0.1$ for the spin-spin autocorrelation function in
 the form $\ln (-\ln C(t))$
against $\ln t$ for both $T > T_G$ and $T < T_G$. Clearly, for each
temperature,
there are two distinct linear regimes: an initial \lq short-time\rq\ linear
regime which becomes more prolonged as the temperature is lowered and a \lq
long-time\rq\ linear regime. In figure 2(a) we also display the lines of best
fit
which were obtained on assuming stretched exponential behaviour (equation (19))
for the \lq long-time\rq\ regimes. The gradients of the plots decrease slowly
with decreasing temperature. Hence, the Kohlrausch law is only satisfied
if we assume a temperature-dependent exponent $n_c(T)$.
\par Table 1 contains values of $n_c(T)$ extracted from figure 2(a) and these
are plotted against $T$ in figure 6 which also shows the values of other
exponents
discussed below.
\par Using the exact expression [17] for $\xi$ for $T > T_G$
and $Z=2.076\pm 0.005$ [18], it
can readily be confirmed that $t\ll \xi^{2+Z}$ for $T=3.0$
for the data
displayed in figure 2(a). (For $T=4.0$ we have $t\sim \xi^{2+Z}$ for the
displayed
data.) This means that we satisfy the
 conditions under which the first part of equation (5) is
expected to hold. It is interesting to note that whereas Bray [1] predicts
$n_c = 2/(2+Z)\approx 0.49$, we find a temperature-dependence that ranges from
$n_c(T=3.0) = 0.723\pm 0.020$ to $n_c(T=1.0)=0.513\pm 0.004$. In our earlier
work [4] on the diluted Ising model exactly at the percolation threshold
we found the Kohlrausch exponent to range from $n_c(T=3.0)=0.704\pm
0.005$ to $n_c(T=1.5)=0.487\pm 0.001$, with $n_c(T=T_G)=0.641\pm 0.006$.
These values are
comparable with those found in the present study below $p_c$.
\par In figure 2(b) we re-plot the data for $T < T_G$ as $\ln (-\ln C(t))$
against $\ln (\ln t)$. This time the exact expression [17] for $\xi$ for $T <
T_G$
implies that the requirement of equation (17) is met. The straight lines, which
are guides to the eye, have gradient 2. It is clear from figure 2(b) that we
have
yet to attain the asymptotic regime.
\par We now turn to our data for the decay of $C(t)$ for $p=0.25$. The data,
which
is displayed in figures 3(a) and 3(b), was analysed in a similar way to that
for
$p=0.1$. Once again, the data in the \lq long-time\rq\ regimes in
 figure 3(a) may be fitted to a stretched-exponential only if we assume a
temperature-dependent exponent $n_c(T)$; values of the exponents extracted from
the fits in figure 3(a) are given in table 2 and displayed in figure 6. The
Kohlrausch exponents would not appear to depend on the concentration, $p$.
\para Figure 3(b) gives a plot of $\ln (-\ln C(t))$ against $\ln (\ln t)$ of
the
data for $T=2.0$. It is rather puzzling to note that the straight
 line of gradient 2 would appear to indicate that
we have already attained the asymptotic limit given by equation (17) even
though
our data do not yet meet the requirement of $t\gg\xi^{2+Z}$ for
 this particular temperature!
\par We now discuss our results for the non-equilibrium decay of the
 magnetization. We plot our data for $M(t)$ for $p=0.1$ in figures 4(a) and
4(b)
which should be compared with figures 2(a) and 2(b), respectively. We see that
the plots are very similar. The stretched-exponential exponents, $n_m(T)$,
extracted from the \lq long-time\rq\ regimes of figure 4(a) are given in
table 1 and displayed in figure 6. We see that the exponents $n_c(T)$ and
$n_m(T)$
cannot be distinguished within the uncertainty of the data, especially for the
higher temperatures. This is to be contrasted with the earlier results [5,7]
which would appear to imply that $n_m(T) > n_c(T)$ for the diluted Ising model
exactly at the percolation threshold.
\par The data for $M(t)$ for $p=0.25$ are presented
 in figures 5(a) and 5(b), respectively.
The values of $n_m(T)$ extracted from figure 5(a) are given in table 2 and
displayed in figure 6. Figure 5(b) is very
 similar to figure 3(b) in that the asymptotic behaviour predicted by equation
(17) is seen even though $t\ll\xi^{2+Z}$ for the data exhibited.
\par We see from figure 6 that the value of the stretched-exponential exponent
appears to depend only on the temperature.
\para 5. CONCLUSION
\par We have presented new Monte Carlo data on the relaxation of the diluted
two-dimensional Ising model below the percolation
 threshold.
\par We find that the non-equilibrium decay of the magnetization and the
relaxation of the equilibrium spin-spin autocorrelation function can both be
described by the Kohlrausch law of a stretched exponential with
a temperature-dependent exponent $n(T)$.
 This result is very similar to previous Monte
Carlo work exactly at the percolation threshold [4-7]. However, unlike the
earlier
work [5-7] for the diluted Ising model exactly at $p_c$,
 we also
 find evidence
that the decay of $M(t)$ and the relaxation of $C(t)$ can be described by
 the same temperature-dependent Kohlrausch exponent.
\par The exact asymptotic behaviour predicted
by a recent theory [3] is not seen even though we managed to obtain values
 of
M(t) and C(t) more than an order of magnitude less than in previous
simulations.
\par It may be that the predicted asymptotic
decay occurs only when $M(t)$ and $C(t)$ are outside any observable regime.
\para 6. ACKNOWLEDGEMENTS
\par Although most of the simulations were performed on the AMT-DAP while
 it was still based at
 Queen Mary and Westfield
College (London University), some additonal runs were made on a Cray YMP at the
Rutherford Appleton Laboratory; the time on the Cray YMP has been made
available
through a research grant (ref: GR/K/00813). The Engineering and
Physical Sciences
 Research Council (EPSRC) of Great Britain is gratefully acknowledged.
\vfill\eject
\para Table 1
\vskip 0.5cm
\vbox{\tabskip=0pt \offinterlineskip
\def\tablerule{\noalign{\hrule}}
\halign to250pt{\strut#& \vrule#\tabskip=1em plus2em&
\hfil#& \vrule#& \hfil#\hfil& \vrule#&
\hfil#& \vrule#\tabskip=0pt\cr\tablerule
&&\hidewidth $T$\hidewidth&&
\hidewidth $n_c$\hidewidth&&
\hidewidth $n_m$\hidewidth&\cr\tablerule
&& 1.0&& $0.513\pm 0.004$&& $0.538\pm 0.004$&\cr\tablerule
&& 1.5&& $0.597\pm 0.012$&& $0.649\pm 0.018$&\cr\tablerule
&& 1.75&& $0.641\pm 0.021$&& $0.661\pm 0.011$&\cr\tablerule
&& 1.90&& $0.686\pm 0.027$&&---\qquad\quad &\cr\tablerule
&& 3.0&& $0.723\pm 0.020$&&---\qquad\quad &\cr\tablerule
&& 4.0&& $0.779\pm 0.017$&& $0.831\pm 0.017$&\cr\tablerule
 \noalign{\smallskip}}}
\vskip 3.0cm
\para Table 2
\vskip 0.5cm
\vbox{\tabskip=0pt \offinterlineskip
\def\tablerule{\noalign{\hrule}}
\halign to250pt{\strut#& \vrule#\tabskip=1em plus2em&
\hfil#& \vrule#& \hfil#\hfil& \vrule#&
\hfil#& \vrule#\tabskip=0pt\cr\tablerule
&&\hidewidth $T$\hidewidth&&
\hidewidth $n_c$\hidewidth&&
\hidewidth $n_m$\hidewidth&\cr\tablerule
&& 2.0&& $0.670\pm 0.013$&& $0.703\pm 0.010$&\cr\tablerule
&& 2.27&& $0.679\pm 0.011$&&---\qquad\quad &\cr\tablerule
&& 3.0&& $0.656\pm 0.029$&& $0.785\pm 0.027$&\cr\tablerule
&& 4.0&& $0.796\pm 0.021$&& $0.827\pm 0.029$&\cr\tablerule
\noalign{\smallskip}}}
\vfill\eject
\para TABLE CAPTIONS
\vskip 1cm
\para Table 1
\para A table of the Kohlrausch exponents $n_c$ and $n_m$ for the decay of the
equilibrium spin-spin auto-correlation function and the non-equilibrium
relaxation of the
magnetization, respectively, for $p=0.1$.
\vskip 1cm
\para Table 2
\para A table of the Kohlrausch exponents $n_c$ and $n_m$ for the decay of the
equilibrium spin-spin auto-correlation function and the non-equilibrium
relaxation of the
magnetization, respectively, for $p=0.25$. Note that $T_G\approx 2.27$.
\vfill\eject
\para FIGURE CAPTIONS
\vskip 1cm
\para Figure 1
\para A schematic phase diagram for a diluted ferromagnet: paramagnetic ($P$),
ferromagnetic ($F$) and Griffiths ($G$) phases are indicated. $T_G$ is the
Griffiths
temperature.
\vskip 0.5cm
\para Figure 2(a)
\para A plot of $\ln (-\ln C(t))$ against $\ln t$ for $p=0.1$. The straight
 lines are best
fits assuming a stretched exponential form for the \lq long-time\rq\ linear
regimes (see text). Figure 6 displays the gradients $n_c(T)$ obtained from
this plot. The temperatures shown
 are:$\circ , T=4.0$; $\square , T=3.0$; $\diamond , T=1.9$; $\triangle ,
T=1.75$;
$\bullet , T=1.5$; and $\square , T=1.0$.
\vskip 0.5cm
\para Figure 2(b)
\para A re-plot of the data displayed in figure 2(a) as $\ln (-\ln C(t))$
against
 $\ln (\ln t)$. The straight lines, which are guides to the eye, have gradient
2. The temperatures shown are:$\diamond , T=1.9$; $\triangle , T=1.75$;
$\bullet , T=1.5$; and $\square , T=1.0$.
\vskip 0.5cm
\para Figure 3(a)
\para A plot of $\ln (-\ln C(t))$ against $\ln t$ for $p=0.25$. The gradients
$n_c(T)$ extracted from the straight line fits are displayed in figure 6.
The temperatures shown are:$\triangle , T=4.0$;
 $\diamond , T=3.0$; $\circ , T=T_G$; and $\square , T=2.0$.
\vskip 0.5cm
\para Figure 3(b)
\para A re-plot of the data for $T=2.0$ shown in figure 3(a) as $\ln (-\ln
C(t))$ against
$\ln (\ln t)$. The straight line has gradient 2. Note that, despite the
reasonable
fit, we have yet to attain the limit $t\gg\xi^{2+Z}$ at this temperature.
\vskip 0.5cm
\para Figure 4(a)
\para A plot of $\ln (-\ln M(t))$ against $\ln t$ for $p=0.1$. The gradients
$n_m(T)$ extracted from the straight line fits over the \lq long-time\rq\
regimes
are shown in figure 6. The temperatures displayed are: $\triangle , T=4.0$;
$\diamond , T=1.75$; $\square , T=1.5$; $\circ , T=1.0$.
\vskip 0.5cm
\para Figure 4(b)
\para A re-plot of some of the data displayed in figure 4(a) as $\ln (-\ln
C(t))$
against $\ln (\ln t)$. The straight lines are guides to the eye and have
gradient
2. The temperatures shown are: $\diamond , T=1.75$; $\square , T=1.5$;
 $\circ , T=1.0$.
\vskip 0.5cm
\para Figure 5(a)
\para A plot of $\ln (-\ln M(t))$ against $\ln t$ for $p=0.25$. The slopes
of the straight lines, $n_m(T)$, are displayed in figure 6. The temperatures
shown
are: $\diamond , T=4.0$; $\square , T=3.0$; $\circ , T=2.0$.
\vskip 0.5cm
\para Figure 5(b)
\para A re-plot of the data for $T=2.0$ displayed in figure 5(a). The straight
line has gradient 2. Note that we are not yet in the limit $t\gg\xi^{2+Z}$;
compare with figure 3(b).
\vskip 0.5cm
\para Figure 6
\para A plot of the Kohlrausch exponents $n_{c,m}(T,p)$
 as extracted from the earlier
figures against the temperature. Symbols: $\circ , n_c(T,0.1)$; $\circ ,
n_m(T,0.1)$;
$\square , n_c(T,0.25)$; $\square , n_m(T,0.25)$.
\vfill\eject
\para REFERENCES
\item {[1]} A J Bray, Phys Rev Lett {\bf 60} 720 (1988)
\item {[2]} D Dhar, M Randeria and J P Sethna, Europhys Lett {\bf 5} 485 (1988)
\item {[3]} A J Bray, J Phys A: Math Gen {\bf 22} L81 (1989)
\item {[4]} S Jain, J Phys C: Solid State Phys {\bf 21} L1045 (1988)
\item {[5]} S G W Colborne and A J Bray, J Phys A: Math Gen {\bf 22} 2505
(1989)
\item {[6]} H Takano and S Miyashita, J Phys Soc Japan {\bf 58} 3871 (1989)
\item {[7]} V B Andreichenko, W Selke and A L Talapov, J Phys A: Math Gen {\bf
25}
 L 283 (1992)
\item {[8]} R B Stinchcombe, {\bf Phase Transitions and Critical Phenomena} vol
7, ed. C Domb and J L Lebowitz, New York: Academic (1983)
\item {[9]} R B Griffiths, Phys Rev Lett {\bf 23} 17 (1969)
\item {[10]} M Randeria, J P Sethna and R G Palmer, Phys Rev Lett {\bf 54} 1321
(1985)
\item {[11]} P C Hohenberg and B I Halperin, Rev Mod Phys {\bf 49} 435 (1972)
\item {[12]} H Kunz and B Souillard, J Stat Phys {\bf 19} 77 (1978)
\item {[13]} A J Bray and G J Rodgers, J Phys C: Solid State Phys {\bf 21}
 L243 (1988)
\item {[14]} R Kohlrausch, Ann Phys, Lpz {\bf 12} 393 (1847)
\item {[15]} S Jain, {\bf Monte Carlo Simulations of Disordered Systems},
 Singapore, World
 Scientific Publishing Co Pte Ltd, (1992)
\item {[16]} S Jain and E J S Lage, J Phys A: Math Gen {\bf 26} L1211 (1993)
\item {[17]} M E Fisher, Rep Prog Phys {\bf 30} 615 (1967)
\item {[18]} M Mori and Y Tsuda, Phys Rev B{\bf 37} 5444 (1988)
\end